\documentclass[prd,superscriptaddress,twocolumn,nofootinbib]{revtex4}

\usepackage{amsfonts,amssymb,amsmath}
\usepackage{color}
\usepackage{graphicx}
\usepackage{dcolumn}
\usepackage{bm}

\newcommand{\BF}[1]{{\mathbf #1}}

\newcommand{\beq}{\begin{equation}}
\newcommand{\eeq}{\end{equation}}
\newcommand{\bea}{\begin{eqnarray}}
\newcommand{\eea}{\end{eqnarray}}
\newcommand{\nn}{\nonumber}

\newcommand{\half}{{1\over 2}}

\newcommand{\Slash}[1]{{\ooalign{\hfil#1\hfil\crcr\raise.167ex\hbox{/}}}}

\begin{document}

\title{A gauge mediation scenario with hidden sector renormalization in MSSM}

\author{Masato Arai}
\email{Masato.Arai(AT)utef.cvut.cz}
\affiliation{Institute of Experimental and Applied Physics,
Czech Technical University in Prague, 
Horsk\' a 3a/22, 128 00 Prague 2, Czech Republic}
\author{Shinsuke Kawai}
\email{kawai(AT)skku.edu}
\affiliation{
Institute for the Early Universe (IEU),
11-1 Daehyun-dong, Seodaemun-gu, Seoul 120-750, Korea} 
\affiliation{Department of Physics, 
Sungkyunkwan University,
Suwon 440-746, Korea}
\author{Nobuchika Okada}
\email{okadan(AT)ua.edu}
\affiliation{
Department of Physics and Astronomy, 
University of Alabama, 
Tuscaloosa, AL35487, USA} 

\date{\today}

\begin{abstract}
We study the hidden sector effects to the mass renormalization of a simplest 
gauge-mediated supersymmetry breaking scenario.
We point out that possible hidden sector contributions render the soft 
scalar masses smaller,
resulting in drastically different sparticle mass spectrum at low energy.
In particular, in the ${\BF 5}+ \bar{\BF 5}$ minimal gauge mediated supersymmetry breaking 
with high messenger scale (that is favored by the gravitino cold dark matter scenario), 
we show that stau can be the next lightest superparticle 
for moderate values of hidden sector self coupling.
This provides a very simple theoretical model of long-lived charged next lightest superparticles, that imply distinctive signals in ongoing and upcoming collider experiments.

\end{abstract}

\maketitle

\section{Introduction}
\label{sec:intro}

The minimal supersymmetric standard model (MSSM) is an attractive framework of 
a theory beyond the standard model.
It resolves some of the shortcomings of the standard model and at the same time
makes falsifiable predictions for upcoming experiments. 
The presence of dark matter, for an instance, has been consolidated by 
numerous cosmological 
and astrophysical observations \cite{Komatsu:2008hk}.
According to the best-fit $\Lambda$-cold dark matter (the `concordance') model,
the major part of the energy density of the present universe is dark energy 
and dark matter, neither 
of them is explained by the standard model.
While dark energy is likely to be of gravitational origin, dark matter is 
readily accommodated in 
MSSM; as the lightest superparticle (LSP) is protected by R-parity, it is an 
ideal candidate of dark matter.

Among competing supersymmetry breaking scenarios of MSSM, 
the gauge-mediated supersymmetry breaking (GMSB) 
\cite{GMSB} 
is advantageous for natural suppression of flavor-changing neutral currents and CP violation.
In GMSB the LSP is gravitino.
Cosmological model building with successful structure formation favors {\em cold} dark matter, which, in the case of gravitino dark matter, yields gravitino mass lower bound 
$m_{3/2}\gtrsim 100$ keV.
Gravitino LSP of this mass range is realized by GMSB models with sufficiently high messenger scale (see also Section 3).
For this reason we will be concerned with the high messenger scale GMSB scenario in this paper.
%
%
The interaction of gravitino with the standard model particles is Planck-suppressed.
While this makes gravitino an ideal candidate of dark matter, it also
makes its direct detection rather difficult.  
A more direct telltale of supersymmetry is from the next lightest 
superparticle (NLSP), which, in the case of GMSB, is either the lightest 
neutralino or the lightest charged slepton(s).
In the high messenger scale GMSB, the NLSP is long-lived, behaving as quasi-stable
particles, and the distinction of whether the NLSP is neutralino or slepton leads to entirely 
different physics.
%
In the `minimal' GMSB scenario based on the $N_5=1$
(the sum of the Dynkin indices) ${\BF 5}+ \bar{\BF 5}$ messengers which
are representations of the $SU(5)$ gauge group \cite{Dimopoulos:1996vz,Dimopoulos:1996yq}, 
the NLSP is neutralino.
The long-lived neutralino NLSP scenario is strongly constrained by the big-bang 
nucleosynthesis (BBN) \cite{Kawasaki:2008qe}, leading to severe constraints on
both gravitino and neutralino masses.
Assuming non-thermal production of gravitino dark matter combined with the BBN bound,
the neutralino mass is constrained to be larger than TeV scale.
In contrast, if the gravitino dark matter is of thermal origin the neutralino mass can be smaller.
In any case, the long-lived neutralino behaves like stable particles in collider experiments
and the physics is similar to that of neutralino LSP.
%
If the NLSP is charged slepton(s), on the other hand, we expect entirely different signals.
A natural candidate for a charged NLSP is the lightest scalar tau (stau).
While the stau NLSP scenario is also constrained by BBN \cite{Kawasaki:2008qe},
the restrictions on the gravitino and stau masses are not so tight as the neutralino NLSP case.
In collider experiments, charged particles like stau leave tracks in detectors, 
allowing direct observations of the NLSP. 
It would also be possible to make a precise measurement of its mass, 
and the absence of missing energy (such as due to neutrino) would facilitate detailed study of 
various processes involving stau 
\cite{Buchmuller:2004rq,Feng:2004gn,Hamaguchi:2004df,
Feng:2004yi,Hamaguchi:2006vu}.
In GMSB, stau NLSP is obtained in models with larger $N_5$
\cite{Dimopoulos:1996vz,Dimopoulos:1996yq}
or messengers belonging to larger gauge groups \cite{Mohapatra:2008wx}.
The messenger index $N_5$, however, cannot be taken to be arbitrarily large 
as it is constrained by the condition that the successful gauge group 
unification must be preserved. 
In the case of simple multiple $SU(5)$ messengers, for example, 
the possible range of $N_5$ is roughly $N_5\lesssim 10$ 
 for the messenger masses $\approx 10^{10}$ GeV. 
For too large $N_5$ the gauge couplings diverge before the unification.

These studies are based on the traditional assumption that the renormalization 
group (RG) flow of the visible world masses does not depend on details of 
the hidden sector dynamics, which, as is realized recently 
\cite{Dine:2004dv,Cohen:2006qc}, turned out not necessarily to be the case. 
Based on this observation, effects of hidden sector on the renormalization 
group equations (RGE) are studied in \cite{Campbell:2008tt}.
They use a toy model of the hidden sector with a self-interaction coupling 
in the constrained MSSM, and show that the soft mass spectrum at low energy is altered by 
hidden sector effects, suggesting that such spectra can be used to determine the hidden scale 
and the strength of the self-coupling.
In the present paper we shall carry out a similar analysis on the GMSB scenario that gives rise to
gravitino cold dark matter, by taking into account the contributions from the hidden sector
and reexamining the RG analysis of the minimal GMSB 
\cite{Dimopoulos:1996vz,Dimopoulos:1996yq}.
We find that the effect of the renormalized self-interaction coupling of the hidden sector field 
renders the soft scalar masses smaller than the conventional GMSB results, 
and consequently the low energy mass spectrum of the MSSM particles can be drastically altered. 
A phenomenologically interesting consequence of our analysis is that stau can be the
NLSP even in the minimal GMSB scenario.
Our model thus provides
a simple stau NLSP model with gravitino cold dark matter.

The plan of the paper is as follows. 
In the next section we present the setup of our model, focusing on how the hidden sector dynamics
contributes to the RG of the visible sector. 
In Section 3 we present our numerical results, and in Section 4 we discuss constraints of the
parameter space from the collider and cosmological experiments. 
We conclude in Section 5 with comments, and explicit forms of the hidden sector contributions 
to the RGE are summarized in the Appendix.

\section{The hidden sector in the gauge-mediated supersymmetry breaking}
\label{sec:model}

In the GMSB scenario, the messengers couple directly to the hidden sector fields
(which is responsible for the supersymmetry breaking), and indirectly 
to the MSSM matter fields through the standard model gauge interactions.
Once the supersymmetry is broken in the hidden sector, 
the gauginos and the scalars in the MSSM become massive 
via one-loop and two-loop corrections, respectively.
In the minimal GMSB based on the $N_5=1$ ${\BF 5}+ \bar{\BF 5}$ 
messenger belonging to an $SU(5)$, 
the gaugino masses at the messenger scale $M$ are given by
\begin{eqnarray}
M_a(t=0)&=&{\alpha_a(M) \over 4\pi}\Lambda,
\label{gaugino-mass}
\end{eqnarray}
where 
\beq
\alpha_a={g_a^2 \over 4\pi}, 
\eeq
$t=\ln(\mu/M)$ and $\Lambda=F/M$ with $F$ the supersymmetry breaking scale.
The index $a=1,2,3$ is for the MSSM gauge groups $U(1)$, $SU(2)$, $SU(3)$, and
$g_a$ are the corresponding gauge couplings.
The soft scalar masses at the messenger scale are given by
\begin{eqnarray}
m_i^2(t=0)&=&2\Lambda^2 \sum_{a=1}^3C_2^a(R_i) \left(\alpha_a(M) \over 4\pi\right)^2, \label{squark-mass}
\end{eqnarray}
where the index $i$ denotes the MSSM scalars
($\tilde{q}, \tilde{u}, \tilde{d}, \tilde{l}, \tilde{e}, H_u, H_d$) 
in this order, 
and $C_2^a(R_i)$ are the quadratic Casimir for the matter fields in 
representation $R_i$ of the $a$-th MSSM gauge group.
As indicated, these masses are generated at the messenger scale $M$.
Physical masses at the electroweak scale are evaluated through 
RG evolution.
Below we follow \cite{Cohen:2006qc,Campbell:2008tt}
and derive the RGE, taking the hidden
sector effects into account.

We model the hidden sector by a single chiral superfield $X$ with Lagrangian, 
\begin{eqnarray}
 {\cal L}_{\rm hid}=\int d^4\theta X^\dagger X + \left(\int d^2\theta {\cal W}+h.c.\right),
\end{eqnarray}
where the superpotential is
\begin{eqnarray}
{\cal W}={\lambda \over 3}X^3. \label{hid}
\end{eqnarray}
This does not itself break the supersymmetry but we assume that the 
$F$-term component of $X$ 
acquires a non-zero vacuum expectation value by some mechanism.
We shall be interested in the effects of the hidden sector renormalization.
The hidden sector field $X$ couples only to the messenger fields, and
the messenger fields couple to the MSSM fields through the gauge interactions.
Below the messenger scale $M$, the messenger fields are 
integrated out, yielding the following effective interactions 
between $X$ and the MSSM fields,
\begin{eqnarray}
 {\cal L}_{\rm int}&=&k_i \int d^4\theta {X X^\dagger \over M^2}\Phi_i \Phi_i^\dagger  \nonumber \\
& &+\left(w_a \int d^2\theta {X \over M}W^{a\alpha} W^a_\alpha + h.c. \right),
\label{h-v}
\end{eqnarray}
with real and complex coefficients $k_i$ and $w_a$.
Here 
$\Phi_i$ are the MSSM matter superfields (their scalar components are 
$\phi_i=\tilde{q}, \tilde{u}, \tilde{d}, \tilde{l}, 
\tilde{e}, H_u, H_d$) and $W_\alpha^a$ are the MSSM field-strengths.
As the $F$-term component of $X$ obtains a non-zero vacuum expectation value at the 
messenger scale $M$, the first and the second terms of (\ref{h-v}) respectively give 
the scalar masses (\ref{squark-mass}) and the gaugino masses (\ref{gaugino-mass}).

Let us consider quantum corrections to (\ref{h-v}).
There are no 1 particle irreducible diagrams that renormalize operators linear in $X$, 
and hence the second line of (\ref{h-v}) receives only the wave function
renormalization coming from the hidden sector field. 
By the renormalization $X\rightarrow Z_X^{-1/2} X$, the second line of (\ref{h-v}) becomes
\begin{eqnarray}
w_a \int d^2\theta Z_X^{-1/2}{X \over M}W^{a\alpha} W^a_\alpha + h.c.
\end{eqnarray} 
The integrand is a holomorphic function whereas the factor $Z_X^{-1/2}$ is real. 
Therefore, due to the non-renormalization theorem \cite{Seiberg:1993vc, Intriligator:1994jr}, 
there is no quantum correction in the second line in (\ref{h-v}).
This means that $w_a$ does not receive any quantum correction at all scales.
On the other hand, the integrand of the first term in (\ref{h-v}) is not a holomorphic
function and therefore the coefficients $k_i$ are renormalized by the visible sector
gauge interactions and the hidden sector coupling $\lambda$ in (\ref{hid}).
They then satisfy the RGE
\begin{eqnarray}
 &&\hspace{-5mm}{d \over dt}k_i(t) =\gamma(t) k_i(t) \nonumber \\ 
 &&\hspace{-5mm}\quad \quad -{1 \over 16\pi^2}\sum_{a=1}^3 8 C_2^a(R_i)g_a^6(t) G_a,~~G_a \equiv w_aw_a^\dagger. 
\label{RGE1}
\end{eqnarray}
The second term is the leading visible sector contribution. 
In the first term $\gamma(t)$ is the anomalous dimension at $t=\ln(\mu/M)$ arising from the 
hidden sector interactions (\ref{hid}).
The anomalous dimension in the lowest order in $\lambda$ is
\begin{eqnarray}
 \gamma(t) = {\lambda(t) \lambda^\dagger(t) \over 2\pi^2},
\end{eqnarray}
where $\lambda(t)$ is the running hidden sector Yukawa coupling which renormalizes according to
\begin{eqnarray}
{d \lambda(t) \over dt}={3 \over 8\pi^2}\lambda^3(t).
\end{eqnarray}
The RGE (\ref{RGE1}) are solved as
\begin{eqnarray}
&& \hspace{-5mm} k_i(t)=\exp\left(-\int_t^0 dt^\prime \gamma(t^\prime)\right)k_i(0) \\
&& \hspace{-5mm} +{1 \over 16\pi^2}\sum_{a=1}^3 8 C_2^a(R_i) \int_t^0d s g_a^6(s) 
 \exp\left(-\int_t^s dt^\prime \gamma(t^\prime) \right) G_a. \nonumber
\end{eqnarray}
Using this expression, the scalar masses of the visible 
sector including the RG effects are described by
\begin{eqnarray}
& \displaystyle m_{i}^2(t)=k_i(t)\Lambda^2, & \label{RGE-s}
\end{eqnarray}
with 
\begin{eqnarray}
 k_i(0) = 2\sum_{a=1}^3\left(\alpha_a(M) \over 4\pi\right)^2C_2^a(R_i).
\end{eqnarray}
Here the gauge couplings run according to the standard 1-loop formula,
\begin{eqnarray}
&&\hspace{-3mm}{1 \over \alpha_a(\mu)} \nn \\
&&\hspace{-3mm}=\left\{
\begin{array}{l}
\displaystyle
{1 \over \alpha_a(M_Z)}-{1 \over 2\pi}b_a \ln{M_S \over M_Z}
\displaystyle
-{1 \over 2\pi}b_a^S \ln{\mu \over M_S}, \\
\hspace{41mm} (\mu > M_S) \\
\displaystyle
{1 \over \alpha_a(M_Z)}-{1 \over 2\pi}b_a \ln{\mu \over M_Z}, \quad (\mu \le M_S)
\end{array}
\right.
\end{eqnarray}
where 
$(b_1^S, b_2^S, b_3^S)=(-3, 1, 33/5)$ for the MSSM
and $(b_1, b_2, b_3)=(-7, -19/6, 41/10)$ for the standard model gauge couplings.
$M_Z$ and $M_{\rm S}$ are the $Z$-boson mass and a typical soft mass scale, 
respectively.

%
\begin{figure}
\begin{center}
\includegraphics[scale=.75]{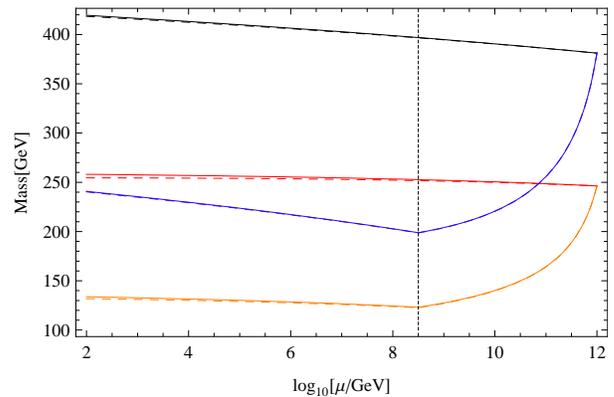}%
\end{center}
\caption{\label{fig_slepton}
The slepton mass RG flows with and without the hidden sector effects.
The upper curves (black and blue) are the left-handed, and the lower curves
(red and orange) are the right-handed sleptons. 
The 1st and the 3rd generations are respectively indicated by the solid and the dashed curves.
The curves with sharp decline (blue and orange) above the hidden scale 
$M_{\rm hid}=10^{8.5}$ GeV 
(indicated by the vertical dotted line) are the flows including the hidden sector effects.
The straight lines (black and red) are the flows without the hidden sector effects.
We have chosen $M=10^{12}$ GeV, $\lambda=3.8$ and $\tan\beta = 10$.
}
\end{figure}
\begin{figure}
\begin{center}
\includegraphics[scale=.75]{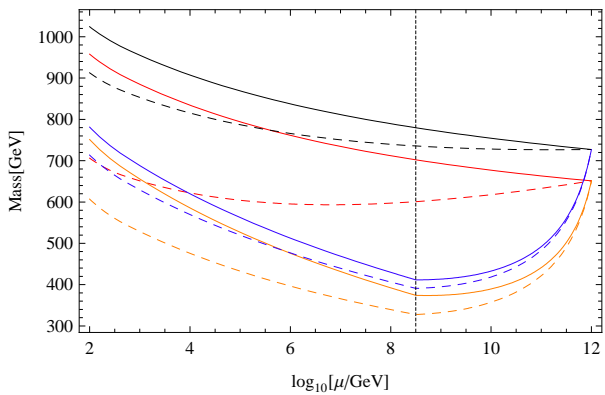}%
\end{center}
\caption{\label{fig_squark}
The squark mass RG flows with and without the hidden sector effects.
The curves starting from the larger mass at $M=10^{12}$ GeV are 
the left-handed squarks, and those starting from the smaller 
mass are the right-handed up-type squarks. 
The 1st and the 3rd generations are denoted by the solid 
and the dashed curves. 
The blue (left-handed) and orange (right-handed) curves 
with sharp decline above the hidden scale 
$M_{\rm hid}=10^{8.5}$ GeV (the dotted vertical line) 
are the flows with the hidden sector effects, 
whereas the black (left-handed) and red (right-handed) curves 
show the flows without the hidden sector effects.
The parameters are the same as in Fig.~\ref{fig_slepton} 
($\lambda=3.8$ and $\tan\beta = 10$).
}
\end{figure}
\begin{figure}
\begin{center}
\includegraphics[scale=.75]{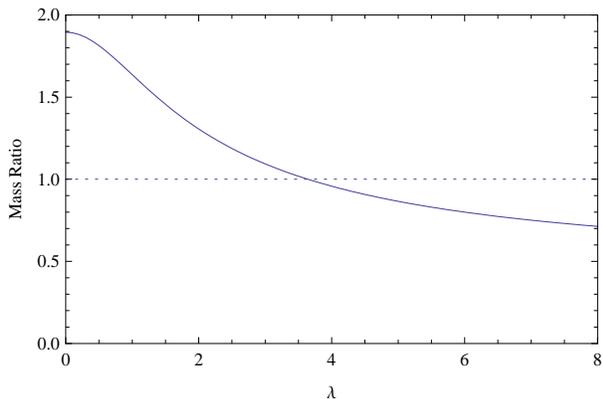}
\end{center}
\caption{\label{fig_lambda}
The stau/neutralino mass ratio 
$m_{\tilde\tau}/m_{\tilde{\chi}^0}$ 
plotted against 
 the hidden sector coupling $\lambda$. 
We used $M=10^{12}$ GeV and $\tan\beta = 10$.
}
\end{figure}

\section{The hidden sector renormalization group flow}

We shall be interested in the low energy mass spectra, 
in particular determination of the NLSP.
The hidden sector renormalization affects the mass RG flows
between the messenger scale $\mu=M$ and the hidden scale $\mu=M_{\rm hid}$.
Eq. (\ref{RGE1}) implies that the RGE for the MSSM soft scalar masses are modified as,
\begin{eqnarray}
 {dm_i^2 \over dt} \rightarrow  {dm_i^2 \over dt}+{\lambda \lambda^\dagger \over 2\pi^2}m_i^2,
 \label{h-eff}
\end{eqnarray}
while those for the other masses and couplings are unaltered 
(see, for example, \cite{Castano:1993ri}). 
Below the hidden scale, the hidden sector fields are integrated out and all
the masses and the couplings evolve according to the standard MSSM RGE.
The low energy mass spectrum is obtained by integrating the RGE, from the messenger
scale to the hidden scale with the hidden sector effects (\ref{h-eff}) included,
and then down to the electroweak scale in the standard way.
In this section we present numerical study of the MSSM RGE including the hidden 
sector effects at one-loop order. 
The non-standard parts of the RGE are listed in Appendix \ref{appendix}.

In our computations we made following approximations.
For simplicity only the $(3,3)$ family component of the three Yukawa matrices,
and only the $(3,3)$ family component of the trilinear A-term matrices, are set to be non-zero.
The latter are set to vanish at the messenger scale. 
Throughout our analysis we use following values:
$\tan\beta=10$, 
the soft mass scale $M_S=500$ GeV, and $\Lambda=F/M=10^5$ GeV.
The hidden sector scale is defined as 
$M_{\rm hid}=\sqrt{F}=\sqrt{\Lambda\times M}$. 
For instance, for $M=10^{12}$ GeV, 
the hidden scale is $M_{\rm hid}=10^{8.5}$ GeV.

Fig.~\ref{fig_slepton} shows the mass RG flows 
of the first and the third generation sleptons
with and without the hidden sector effects, 
and Fig.~\ref{fig_squark} shows similar results for the squarks.
The messenger scale is taken to be $M=10^{12}$ GeV.
In both Fig.~\ref{fig_slepton} and Fig.~\ref{fig_squark}, 
we see that the hidden sector renormalization
(down to the hidden scale $M_{\rm hid}$) gives rise 
to smaller soft scalar masses, compared to the flows without it. 
This effect is readily understood by looking at the RGE
(\ref{eqn:RGE_Q})-(\ref{eqn:RGE_Hd}), in which the effects of the hidden sector renormalization are the positive terms $4\lambda^2m_i^2$ on the right hand sides\footnote{
In GMSB a hidden sector field need be an MSSM gauge singlet and a possible model
of hidden sector superpotential is Yukawa type, as (\ref{hid}).
Then the hidden sector contributions to RGE are always positive.
}.
It is also obvious from the RGE that larger $\lambda$ gives smaller soft scalar masses at low
energy.
In contrast, the RGE and hence the low energy mass spectrum of the 
gauginos are not affected by the hidden sector renormalization.
This implies that the lightest scalar superparticle, namely the lightest stau, can be lighter than the 
lightest gaugino, i.e. bino, for sufficiently large $\lambda$.
In Fig.~\ref{fig_lambda}, we plot the stau/neutralino mass ratio 
$m_{\tilde{\tau}}/m_{\tilde{\chi}^0}$ at the electroweak scale
as a function of $\lambda$ with $M=10^{12}$ GeV.
$\lambda=0$ corresponds to no hidden sector RG effects, reproducing the known result of
neutralino NLSP in the minimal GMSB scenario.
As $\lambda$ is increased\footnote{
The perturbative means is valid for $\lambda^2/(4\pi)\lesssim 1$.
Our choice of the parameter in the following discussions is roughly within this range.
}, the mass $m_{\tilde\tau}$ of the lightest stau 
$\tilde\tau$
becomes smaller, while the neutralino mass $m_{\tilde{\chi}^0}$ remains the same;
the stau becomes lighter than the neutralino when $\lambda\gtrsim 3.7$.
For example, $M=10^{12}$ GeV and $\lambda = 3.8$ yields
$m_{\tilde\tau} = 132$ GeV and the neutralino (bino) mass 
$m_{\tilde{\chi}^0}=135$ GeV at the electroweak scale.
This stau mass is compatible with the LEP bound for long-lived massive charged
particles $\gtrsim$ 102.0 GeV \cite{Abbiendi:2003yd}.

\begin{figure}
\begin{center}
\includegraphics[scale=.75]{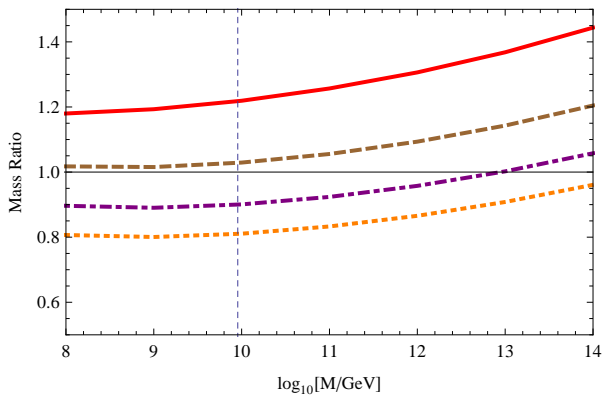}
\end{center}
\caption{\label{fig_RG}
The stau/neutralino mass ratio $m_{\tilde\tau}/m_{\tilde{\chi}^0}$ 
plotted against the messenger scale $M$, 
for $\lambda=2$ (solid red), $\lambda=3$ (dashed brown), 
$\lambda=4$ (dot-dashed purple), $\lambda=5$ (dotted orange) from above, 
with $\tan\beta = 10$.
The vertical dashed line indicates the minimal messenger scale in
our model, $M_*= 9.0\times 10^9$ GeV.
For large enough $\lambda$ stau becomes lighter than neutralino.
}
\end{figure}

The ratio of the stau and the bino masses $m_{\tilde{\tau}}/m_{\tilde{\chi}^0}$ 
is depicted in Fig.~\ref{fig_RG}, against the messenger scale $M$
for various values of $\lambda$. 
The vertical dashed line indicates the minimal messenger scale 
determined by the condition that gravitino is cold dark matter, 
$m_{3/2}\gtrsim 0.1$ MeV.
Here the gravitino mass is given by \cite{Deser:1977uq,Cremmer:1978iv}
\begin{eqnarray}
 m_{3/2}\sim {F \over \sqrt{3}M_{P}}={M \over \sqrt{3}M_P}\Lambda,
\label{gravitino}
\end{eqnarray}
where $M_P=2.4\times 10^{18}$ GeV is the reduced Planck mass.
The gravitino mass is also affected by the hidden sector renormalization 
\cite{Campbell:2008tt}.
For example, when the source of the supersymmetry breaking is a term linear in $X$, 
the vacuum expectation value of the $F$-term component 
of $X$ is proportional to the wave function renormalization factor
$Z_X^{-1/2}=\exp(-\frac{1}{2}\int_t^0 dt \gamma(t))$.
This takes effect from the messenger scale down to the hidden scale, resulting in
suppression of the gravitino mass by the factor $Z_{X}^{-1/2}\approx 0.5$ for $\lambda=2\sim 4$.
Taking this suppression into account, 
the minimal messenger scale consistent with the cold dark matter condition 
$m_{3/2}\gtrsim 0.1$ MeV 
is $M_*\approx 9.0 \times 10^9$ GeV. 
By choosing an appropriate messenger scale $M$ within
$M_*\lesssim M\lesssim10^{12}$ GeV,
stau can be the NLSP with $\lambda$ as small as $3.0$.

\section{Stau NLSP phenomenology}

In the previous section we have seen that the hidden sector renormalization
can change the low energy mass spectrum. 
Here we describe phenomenological implications when the 
NLSP is the lightest stau.

\begin{figure}
\begin{center}
\includegraphics[scale=.94]{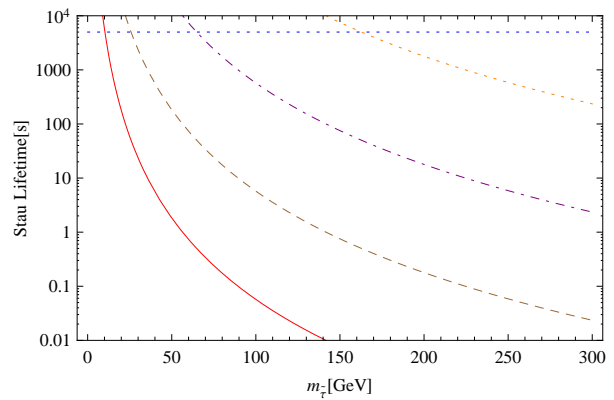}
\end{center}
\caption{\label{fig_decay}
Lifetime of stau plotted against the stau mass, 
for $m_{3/2}=1$ MeV (red solid curve), $m_{3/2}=10$ MeV (brown dashed),
$m_{3/2}=0.1$ GeV (purple dot-dashed) and $m_{3/2}=1$ GeV (orange dotted).
}
\end{figure}

We note that cosmological constraints on the mass parameters in the stau NLSP scenario  
are less stringent than in the neutralino NLSP case \cite{Kawasaki:2008qe}.
In BBN, stau produces a bound state with ${}^4$He that tends to
overproduce primordial ${}^6$Li through catalytic reaction 
${}^4{\rm He}\ \tilde\tau^- +{\rm D}\rightarrow {}^6{\rm Li}+\tilde\tau^-$
\cite{Pospelov:2006sc,Cyburt:2006uv} 
(see also \cite{Kohri:2006cn,Kaplinghat:2006qr,Kawasaki:2008qe}),
giving upper bound for the stau lifetime\footnote{
This constraint may be relaxed by considering substantial leftÐright mixing of the stau mass 
eigenstates \cite{Ratz:2008qh}.}
\bea
\tau_{\tilde\tau}&=&\Gamma^{-1}(\tilde\tau\rightarrow\tau\tilde G)
\simeq\frac{48\pi m_{3/2}^2M_P^2}{m_{\tilde\tau}^5}
\left(1-\frac{m_{3/2}^2}{m_{\tilde\tau}^2}\right)^{-4}\nn\\ 
&\simeq & 5.7 \times 10^4 
\left( \frac{m_{3/2}}{1\;{\rm GeV}}  \right)^2 
\left( \frac{100 \; {\rm GeV}}{m_{\tilde{\tau}}}  \right)^5 \nn\\
&\lesssim& 5\times 10^3 
\ {\rm seconds}.
\eea
Fig.~\ref{fig_decay} shows the stau lifetime $\tau_{\tilde\tau}$ 
plotted against the stau mass
$m_{\tilde\tau}$ for different values of gravitino mass $m_{3/2}$.
The upper bound of the stau lifetime $\tau_{\tilde\tau}\lesssim 5\times 10^3$ constrains the 
gravitino mass from above. 
For our typical range of stau mass $m_{\tilde\tau}\approx 130$ GeV, the gravitino
mass is $m_{3/2}\lesssim 0.6$ GeV. 
Together with the condition that gravitino behaves as cold dark matter,
a successful cosmological scenario with stau NLSP and 
gravitino cold dark matter requires the gravitino
mass to be within $0.1$ MeV $\lesssim m_{3/2} \lesssim 0.6$ GeV.
The model prediction (\ref{gravitino}) is safely within this range.

The origin of relic gravitino is either thermal or non-thermal productions.
The thermal production is due to scattering in the thermal plasma in the radiation
dominated era after inflation. 
The relic abundance from this contribution is computed as 
\cite{Bolz:2000fu,Pradler:2006qh,Steffen:2006hw}
\beq
\Omega_{\tilde G}^{\rm TP}h^2
\simeq 
0.3
\left(\frac{T_R}{10^{10} {\rm GeV}}\right)
\left(\frac{100 {\rm GeV}}{m_{3/2}}\right)
\left(\frac{M_3}{1 {\rm TeV}}\right)^2,
\label{eqn:ThAb}
\eeq
where $T_R$ is the reheating temperature and $h$ is the Hubble constant in the unit of
100 km Mpc${}^{-1}$s${}^{-1}$.
The non-thermal production of gravitino is due to decay of the NLSP
\footnote{
The productions of gravitino from decay of moduli, inflation, and Polonyi fields have also been
discussed \cite{Hashimoto:1998mu,Kallosh:1999jj,Felder:1999pv,Maroto:1999ch,Giudice:1999yt,Dine:1983ys,Coughlan:1984yk,Joichi:1994ce,Heckman:2008jy}.} 
\cite{Feng:2003xh,Feng:2003uy}.
The abundance of non-thermally produced gravitino is estimated as
\cite{Borgani:1996ag,Asaka:2000zh,Steffen:2006wx,Steffen:2007sp}
\beq
\Omega_{\tilde G}^{\rm NTP}h^2
=\frac{m_{3/2}}{m_{\tilde\tau}}\Omega_{\tilde\tau}^{\rm th}h^2
\simeq 0.02\left(\frac{m_{3/2}}{100 {\rm GeV}}\right)
\left(\frac{m_{\tilde\tau}}{1 {\rm TeV}}\right),
\label{eqn:NonThAb}
\eeq
where $\Omega_{\tilde\tau}^{\rm th}$ is the thermal abundance of stau.
The total gravitino abundance is thus
\beq
\Omega_{\tilde G}h^2=\Omega_{\tilde G}^{\rm TP}h^2+\Omega_{\tilde G}^{\rm NTP}h^2,
\eeq
which is constrained by the observed dark matter density \cite{Komatsu:2008hk}
\beq
\Omega_{\tilde G} h^2\leq\Omega_{CDM}h^2\simeq 0.1131\pm0.0034.
\label{eqn:OmegaCDM}
\eeq

For stau mass $m_{\tilde\tau}\approx $ 130 GeV, 
the BBN bound for the gravitino mass $m_{3/2}\lesssim 0.6$ GeV 
restricts the non-thermally produced gravitino abundance 
(\ref{eqn:NonThAb}) to be extremely small, 
$\Omega_{\tilde G}^{\rm NTP}h^2\lesssim 10^{-5}$, indicating that most of the gravitino
dark matter is produced thermally. 
The BBN bound for the gravitino mass and the dark matter abundance 
(\ref{eqn:OmegaCDM}) gives an upper bound of the reheating temperature.
Using (\ref{eqn:ThAb}), we find $T_{R}\lesssim 10^{7}$ GeV.
The model prediction (\ref{gravitino}) with $M=10^{12}$ GeV and $\Lambda=10^5$ GeV
yields $m_{3/2}\approx 11$ MeV; if the right amount of gravitino dark matter of this mass
is to be produced thermally ($\Omega_{CDM}=\Omega_{\tilde G}^{\rm TP}$), one needs 
$T_R\approx 10^{6}$ GeV.
This is lower than the reheating temperature required for successful leptogenesis; the 
observed baryon asymmetry needs to be generated by some other mechanism, such as 
\cite{Affleck:1984fy}.

Stau NLSP is also of interest for collider phenomenology. 
Eq. (17) implies that the lifetime of stau $\tau_{\tilde{\tau}}$ may be long,
depending on the gravitino and stau masses; for instance,
$\tau_{\tilde{\tau}}\approx 100$ sec for $m_{\tilde{\tau}}=130$ GeV and
$m_{3/2}=0.1$ GeV. 
For a particle with such a long lifetime, the decay length well exceeds the size of detectors
and the decay takes place outside detectors. 
There have been intriguing proposals for trapping such long-lived charged NLSPs outside
detectors
\cite{Buchmuller:2004rq,Feng:2004gn,Hamaguchi:2004df,
Feng:2004yi,Hamaguchi:2006vu}.
It has been pointed out that detailed studies of stau decay may provide precise
measurements of the gravitino mass and the supersymmetry breaking scale. 
This would also be a crucial test of supergravity since the stau decay
$\tilde{\tau}\rightarrow \tau \tilde{G}$ involves supergravity effects.

\section{Discussion}

In this paper we studied the effects of hidden sector renormalization 
in the minimal GMSB model, using a toy model of the hidden 
sector field $X$ with self-coupling $\lambda$.
We saw that the soft scalar and gravitino masses are 
affected by the hidden sector RG;
they become smaller than the conventional GMSB RG results.
Since the gaugino masses are unaltered, 
the resulting low energy mass spectrum of the superparticles can be 
quite different from the conventional GMSB.
In particular, we find that stau, instead of neutralino, can be the NLSP in the minimal GMSB model.
Based on this observation we also discussed phenomenological implications
of stau NLSP and gravitino cold dark matter arising from the model.
With our typical choice of parameter values: the messenger scale $M=10^{12}$ GeV, $\tan\beta=10$, $\Lambda=F/M=10^5$ GeV
and the hidden sector coupling $\lambda=3.8$, the model yields stau NLSP with mass
$m_{\tilde\tau}\approx 130$ GeV.
The gravitino mass arising from these parameter values is shown to be consistent with the
BBN constraints. 
It was also shown that gravitino cold dark matter with observationally consistent abundance
is attributed to thermal production in this scenario.

We comment that the hidden sector dependence of the mass spectra, as discussed in this paper,
does not necessarily mar the predictability of the GMSB scenario.
Rather, the hidden sector only influences the soft scalar and the gravitino masses, and the beauty of the GMSB
scenario is kept intact. 
In fact, as emphasized in \cite{Dine:2004dv,Cohen:2006qc,Campbell:2008tt},
the effects of the hidden sector RG can be viewed as a new window to look into the 
hidden sector using the low energy mass spectrum that is expected to be revealed in 
experiments such as Large Hadron Collider. 
While the model discussed here represents a simple toy example of many possible hidden sector
models, we believe essential features of our results to be generic. 
For example, only the soft scalar and the gravitino masses, and not the 
other masses neither the couplings, are subject to the hidden sector RG 
effects. 
This is a consequence of the powerful non-renormalization theorem and we do 
not expect this feature to be dependent on details of the hidden sector. 
Needless to say, however, there might be various other model-dependent issues  
that deserve further study.

\hspace{10mm}

\subsection*{Acknowledgments}
This work was supported in part by the Research Program MSM6840770029 
and by the project of International Cooperation ATLAS-CERN 
of the Ministry of Education, Youth and Sports of 
the Czech Republic (M.~A.), and 
the WCU grant R32-2008-000-10130-0 (S.~K.). 

\bigskip
\begin{appendix}
\section{The renormalization group equations}\label{appendix}

In this appendix we list the RGE for the soft mass parameters 
that involve the hidden sector contributions.
The new ingredient is the hidden-visible interaction (\ref{hid}), 
which give additional terms to the soft mass RGE.
Within our approximation only the $(3,3)$ family component of the three 
Yukawa matrices, $y_t, y_b, y_\tau$, are set to be non-zero.
The corresponding non-zero components of the three trilinear A-term 
matrices are $a_t, a_b, a_\tau$.
We use normalized variables
$A_t=a_t/y_t$, $A_b=a_b/y_b$, $A_\tau=a_\tau/y_\tau$.

The RGE for the soft masses are
\begin{widetext}
\bea
8\pi^2\frac{d m^2_Q}{dt}&=&\xi_t+\xi_b-\frac{16}{3}g_3^2 M_3^2-3g_2^2M_2^2
-\frac{1}{15}g_1^2 M_1^2
+\frac 15 g_1^2 \xi_1+4\lambda^2 m_Q^2,
\label{eqn:RGE_Q}\\
8\pi^2\frac{d m^2_U}{dt}&=&2\xi_t-\frac{16}{3}g_3^2 M_3^2-\frac{16}{15}g_1^2 M_1^2
-\frac 45 g_1^2\xi_1+4\lambda^2 m_U^2,
\eea
\bea
8\pi^2\frac{d m^2_D}{dt}&=&2\xi_b-\frac{16}{3}g_3^2 M_3^2-\frac{4}{15}g_1^2 M_1^2
+\frac 25 g_1^2\xi_1+4\lambda^2 m_D^2,\\
8\pi^2\frac{d m^2_L}{dt}&=&\xi_\tau-3 g_2^2 M_2^2-\frac{3}{5}g_1^2 M_1^2
-\frac 35 g_1^2 \xi_1+4\lambda^2 m_L^2,\\
8\pi^2\frac{d m^2_E}{dt}&=&2\xi_\tau-\frac{12}{5}g_1^2 M_1^2+\frac 65 g_1^2\xi_1
+4\lambda^2 m_E^2,\\
8\pi^2\frac{d m^2_{H_u}}{dt}&=&3\xi_t-3g_2^2 M_2^2-\frac{3}{5}g_1^2 M_1^2
+\frac 35 g_1^2\xi_1+4\lambda^2 m_{H_u}^2,\\
8\pi^2\frac{d m^2_{H_d}}{dt}&=&3\xi_b+\xi_\tau-3 g_2^2 M_2^2-\frac{3}{5}g_1^2 M_1^2
-\frac 35 g_1^2 \xi_1+4\lambda^2 m_{H_d}^2
\label{eqn:RGE_Hd},
\eea
\end{widetext}
where
\bea
\xi_t&=&y_t^2(m_{H_u}^2+m_Q^2+m_U^2+A_t^2),\\
\xi_b&=&y_b^2(m_{H_d}^2+m_Q^2+m_D^2+A_b^2),\\
\xi_\tau&=&y_\tau^2(m_{H_d}^2+m_L^2+m_E^2+A_\tau^2),
\eea
and
\bea
\xi_1&=&\half\left\{m_{H_u}^2 -m_{H_d}^2 \right. \nn \\
&+&\left. {\rm Tr}(m_Q^2-2m_U^2+m_D^2+m_E^2-m_L^2)\right\}.
\eea
Here the trace means the sum over the generations. 
The above equations for the squarks and sleptons apply to the 3rd family; the equations for the 
1st and the 2nd families are obtained from above by simply omitting the Yukawa and the A-term contributions (i.e. neglecting the $\xi_t$, $\xi_b$, $\xi_\tau$ terms).
All the other RGE (the gauge couplings, 
the gaugino masses, the Yukawa couplings and the A-terms) 
are unaffected and are given e.g. in \cite{Castano:1993ri}. 

\end{appendix}


\begin{thebibliography}{51}
\expandafter\ifx\csname natexlab\endcsname\relax\def\natexlab#1{#1}\fi
\expandafter\ifx\csname bibnamefont\endcsname\relax
  \def\bibnamefont#1{#1}\fi
\expandafter\ifx\csname bibfnamefont\endcsname\relax
  \def\bibfnamefont#1{#1}\fi
\expandafter\ifx\csname citenamefont\endcsname\relax
  \def\citenamefont#1{#1}\fi
\expandafter\ifx\csname url\endcsname\relax
  \def\url#1{\texttt{#1}}\fi
\expandafter\ifx\csname urlprefix\endcsname\relax\def\urlprefix{URL }\fi
\providecommand{\bibinfo}[2]{#2}
\providecommand{\eprint}[2][]{\url{#2}}

\bibitem[{\citenamefont{Komatsu et~al.}(2009)}]{Komatsu:2008hk}
\bibinfo{author}{\bibfnamefont{E.}~\bibnamefont{Komatsu}} \bibnamefont{et~al.}
  (\bibinfo{collaboration}{WMAP}), \bibinfo{journal}{Astrophys. J. Suppl.}
  \textbf{\bibinfo{volume}{180}}, \bibinfo{pages}{330} (\bibinfo{year}{2009}),
  \eprint{0803.0547}.

\bibitem{GMSB}
\bibinfo{author}{\bibfnamefont{M.}~\bibnamefont{Dine}},
  \bibinfo{author}{\bibfnamefont{W.}~\bibnamefont{Fischler}}, \bibnamefont{and}
  \bibinfo{author}{\bibfnamefont{M.}~\bibnamefont{Srednicki}},
  \bibinfo{journal}{Nucl. Phys.} \textbf{\bibinfo{volume}{B189}},
  \bibinfo{pages}{575} (\bibinfo{year}{1981});
%
\bibinfo{author}{\bibfnamefont{S.}~\bibnamefont{Dimopoulos}} \bibnamefont{and}
  \bibinfo{author}{\bibfnamefont{S.}~\bibnamefont{Raby}},
  \bibinfo{journal}{Nucl. Phys.} \textbf{\bibinfo{volume}{B192}},
  \bibinfo{pages}{353} (\bibinfo{year}{1981});
%
\bibinfo{author}{\bibfnamefont{L.}~\bibnamefont{Alvarez-Gaume}},
  \bibinfo{author}{\bibfnamefont{M.}~\bibnamefont{Claudson}}, \bibnamefont{and}
  \bibinfo{author}{\bibfnamefont{M.~B.} \bibnamefont{Wise}},
  \bibinfo{journal}{Nucl. Phys.} \textbf{\bibinfo{volume}{B207}},
  \bibinfo{pages}{96} (\bibinfo{year}{1982});
%
\bibinfo{author}{\bibfnamefont{M.}~\bibnamefont{Dine}} \bibnamefont{and}
  \bibinfo{author}{\bibfnamefont{W.}~\bibnamefont{Fischler}},
  \bibinfo{journal}{Nucl. Phys.} \textbf{\bibinfo{volume}{B204}},
  \bibinfo{pages}{346} (\bibinfo{year}{1982});
%
\bibinfo{author}{\bibfnamefont{C.~R.} \bibnamefont{Nappi}} \bibnamefont{and}
  \bibinfo{author}{\bibfnamefont{B.~A.} \bibnamefont{Ovrut}},
  \bibinfo{journal}{Phys. Lett.} \textbf{\bibinfo{volume}{B113}},
  \bibinfo{pages}{175} (\bibinfo{year}{1982});
%
\bibinfo{author}{\bibfnamefont{M.}~\bibnamefont{Dine}} \bibnamefont{and}
  \bibinfo{author}{\bibfnamefont{A.~E.} \bibnamefont{Nelson}},
  \bibinfo{journal}{Phys. Rev.} \textbf{\bibinfo{volume}{D48}},
  \bibinfo{pages}{1277} (\bibinfo{year}{1993}), \eprint{hep-ph/9303230};
%
\bibinfo{author}{\bibfnamefont{M.}~\bibnamefont{Dine}},
  \bibinfo{author}{\bibfnamefont{A.~E.} \bibnamefont{Nelson}},
  \bibnamefont{and} \bibinfo{author}{\bibfnamefont{Y.}~\bibnamefont{Shirman}},
  \bibinfo{journal}{Phys. Rev.} \textbf{\bibinfo{volume}{D51}},
  \bibinfo{pages}{1362} (\bibinfo{year}{1995}), \eprint{hep-ph/9408384};
%
\bibinfo{author}{\bibfnamefont{M.}~\bibnamefont{Dine}},
  \bibinfo{author}{\bibfnamefont{A.~E.} \bibnamefont{Nelson}},
  \bibinfo{author}{\bibfnamefont{Y.}~\bibnamefont{Nir}}, \bibnamefont{and}
  \bibinfo{author}{\bibfnamefont{Y.}~\bibnamefont{Shirman}},
  \bibinfo{journal}{Phys. Rev.} \textbf{\bibinfo{volume}{D53}},
  \bibinfo{pages}{2658} (\bibinfo{year}{1996}), \eprint{hep-ph/9507378}.

\bibitem[{\citenamefont{Dimopoulos et~al.}(1996)\citenamefont{Dimopoulos, Dine,
  Raby, and Thomas}}]{Dimopoulos:1996vz}
\bibinfo{author}{\bibfnamefont{S.}~\bibnamefont{Dimopoulos}},
  \bibinfo{author}{\bibfnamefont{M.}~\bibnamefont{Dine}},
  \bibinfo{author}{\bibfnamefont{S.}~\bibnamefont{Raby}}, \bibnamefont{and}
  \bibinfo{author}{\bibfnamefont{S.~D.} \bibnamefont{Thomas}},
  \bibinfo{journal}{Phys. Rev. Lett.} \textbf{\bibinfo{volume}{76}},
  \bibinfo{pages}{3494} (\bibinfo{year}{1996}), \eprint{hep-ph/9601367}.

\bibitem[{\citenamefont{Dimopoulos et~al.}(1997)\citenamefont{Dimopoulos,
  Thomas, and Wells}}]{Dimopoulos:1996yq}
\bibinfo{author}{\bibfnamefont{S.}~\bibnamefont{Dimopoulos}},
  \bibinfo{author}{\bibfnamefont{S.~D.} \bibnamefont{Thomas}},
  \bibnamefont{and} \bibinfo{author}{\bibfnamefont{J.~D.} \bibnamefont{Wells}},
  \bibinfo{journal}{Nucl. Phys.} \textbf{\bibinfo{volume}{B488}},
  \bibinfo{pages}{39} (\bibinfo{year}{1997}), \eprint{hep-ph/9609434}.

\bibitem[{\citenamefont{Kawasaki et~al.}(2008)\citenamefont{Kawasaki, Kohri,
  Moroi, and Yotsuyanagi}}]{Kawasaki:2008qe}
\bibinfo{author}{\bibfnamefont{M.}~\bibnamefont{Kawasaki}},
  \bibinfo{author}{\bibfnamefont{K.}~\bibnamefont{Kohri}},
  \bibinfo{author}{\bibfnamefont{T.}~\bibnamefont{Moroi}}, \bibnamefont{and}
  \bibinfo{author}{\bibfnamefont{A.}~\bibnamefont{Yotsuyanagi}},
  \bibinfo{journal}{Phys. Rev.} \textbf{\bibinfo{volume}{D78}},
  \bibinfo{pages}{065011} (\bibinfo{year}{2008}), \eprint{0804.3745}.

\bibitem[{\citenamefont{Buchmuller et~al.}(2004)\citenamefont{Buchmuller,
  Hamaguchi, Ratz, and Yanagida}}]{Buchmuller:2004rq}
\bibinfo{author}{\bibfnamefont{W.}~\bibnamefont{Buchmuller}},
  \bibinfo{author}{\bibfnamefont{K.}~\bibnamefont{Hamaguchi}},
  \bibinfo{author}{\bibfnamefont{M.}~\bibnamefont{Ratz}}, \bibnamefont{and}
  \bibinfo{author}{\bibfnamefont{T.}~\bibnamefont{Yanagida}},
  \bibinfo{journal}{Phys. Lett.} \textbf{\bibinfo{volume}{B588}},
  \bibinfo{pages}{90} (\bibinfo{year}{2004}), \eprint{hep-ph/0402179}.

\bibitem[{\citenamefont{Feng et~al.}(2004)\citenamefont{Feng, Rajaraman, and
  Takayama}}]{Feng:2004gn}
\bibinfo{author}{\bibfnamefont{J.~L.} \bibnamefont{Feng}},
  \bibinfo{author}{\bibfnamefont{A.}~\bibnamefont{Rajaraman}},
  \bibnamefont{and} \bibinfo{author}{\bibfnamefont{F.}~\bibnamefont{Takayama}},
  \bibinfo{journal}{Int. J. Mod. Phys.} \textbf{\bibinfo{volume}{D13}},
  \bibinfo{pages}{2355} (\bibinfo{year}{2004}), \eprint{hep-th/0405248}.

\bibitem[{\citenamefont{Hamaguchi et~al.}(2004)\citenamefont{Hamaguchi, Kuno,
  Nakaya, and Nojiri}}]{Hamaguchi:2004df}
\bibinfo{author}{\bibfnamefont{K.}~\bibnamefont{Hamaguchi}},
  \bibinfo{author}{\bibfnamefont{Y.}~\bibnamefont{Kuno}},
  \bibinfo{author}{\bibfnamefont{T.}~\bibnamefont{Nakaya}}, \bibnamefont{and}
  \bibinfo{author}{\bibfnamefont{M.~M.} \bibnamefont{Nojiri}},
  \bibinfo{journal}{Phys. Rev.} \textbf{\bibinfo{volume}{D70}},
  \bibinfo{pages}{115007} (\bibinfo{year}{2004}), \eprint{hep-ph/0409248}.

\bibitem[{\citenamefont{Feng and Smith}(2005)}]{Feng:2004yi}
\bibinfo{author}{\bibfnamefont{J.~L.} \bibnamefont{Feng}} \bibnamefont{and}
  \bibinfo{author}{\bibfnamefont{B.~T.} \bibnamefont{Smith}},
  \bibinfo{journal}{Phys. Rev.} \textbf{\bibinfo{volume}{D71}},
  \bibinfo{pages}{015004} (\bibinfo{year}{2005}), \eprint{hep-ph/0409278}.

\bibitem[{\citenamefont{Hamaguchi et~al.}(2007)\citenamefont{Hamaguchi, Nojiri,
  and de~Roeck}}]{Hamaguchi:2006vu}
\bibinfo{author}{\bibfnamefont{K.}~\bibnamefont{Hamaguchi}},
  \bibinfo{author}{\bibfnamefont{M.~M.} \bibnamefont{Nojiri}},
  \bibnamefont{and} \bibinfo{author}{\bibfnamefont{A.}~\bibnamefont{de~Roeck}},
  \bibinfo{journal}{JHEP} \textbf{\bibinfo{volume}{03}}, \bibinfo{pages}{046}
  (\bibinfo{year}{2007}), \eprint{hep-ph/0612060}.

\bibitem[{\citenamefont{Mohapatra et~al.}(2008)\citenamefont{Mohapatra, Okada,
  and Yu}}]{Mohapatra:2008wx}
\bibinfo{author}{\bibfnamefont{R.~N.} \bibnamefont{Mohapatra}},
  \bibinfo{author}{\bibfnamefont{N.}~\bibnamefont{Okada}}, \bibnamefont{and}
  \bibinfo{author}{\bibfnamefont{H.-B.} \bibnamefont{Yu}},
  \bibinfo{journal}{Phys. Rev.} \textbf{\bibinfo{volume}{D78}},
  \bibinfo{pages}{075011} (\bibinfo{year}{2008}), \eprint{0807.4524}.

\bibitem[{\citenamefont{Dine et~al.}(2004)}]{Dine:2004dv}
\bibinfo{author}{\bibfnamefont{M.}~\bibnamefont{Dine}} \bibnamefont{et~al.},
  \bibinfo{journal}{Phys. Rev.} \textbf{\bibinfo{volume}{D70}},
  \bibinfo{pages}{045023} (\bibinfo{year}{2004}), \eprint{hep-ph/0405159}.

\bibitem[{\citenamefont{Cohen et~al.}(2007)\citenamefont{Cohen, Roy, and
  Schmaltz}}]{Cohen:2006qc}
\bibinfo{author}{\bibfnamefont{A.~G.} \bibnamefont{Cohen}},
  \bibinfo{author}{\bibfnamefont{T.~S.} \bibnamefont{Roy}}, \bibnamefont{and}
  \bibinfo{author}{\bibfnamefont{M.}~\bibnamefont{Schmaltz}},
  \bibinfo{journal}{JHEP} \textbf{\bibinfo{volume}{02}}, \bibinfo{pages}{027}
  (\bibinfo{year}{2007}), \eprint{hep-ph/0612100}.

\bibitem[{\citenamefont{Campbell et~al.}(2008)\citenamefont{Campbell, Ellis,
  and Maybury}}]{Campbell:2008tt}
\bibinfo{author}{\bibfnamefont{B.~A.} \bibnamefont{Campbell}},
  \bibinfo{author}{\bibfnamefont{J.}~\bibnamefont{Ellis}}, \bibnamefont{and}
  \bibinfo{author}{\bibfnamefont{D.~W.} \bibnamefont{Maybury}}
  (\bibinfo{year}{2008}), \eprint{0810.4877}.

\bibitem[{\citenamefont{Seiberg}(1993)}]{Seiberg:1993vc}
\bibinfo{author}{\bibfnamefont{N.}~\bibnamefont{Seiberg}},
  \bibinfo{journal}{Phys. Lett.} \textbf{\bibinfo{volume}{B318}},
  \bibinfo{pages}{469} (\bibinfo{year}{1993}), \eprint{hep-ph/9309335}.

\bibitem[{\citenamefont{Intriligator et~al.}(1994)\citenamefont{Intriligator,
  Leigh, and Seiberg}}]{Intriligator:1994jr}
\bibinfo{author}{\bibfnamefont{K.~A.} \bibnamefont{Intriligator}},
  \bibinfo{author}{\bibfnamefont{R.~G.} \bibnamefont{Leigh}}, \bibnamefont{and}
  \bibinfo{author}{\bibfnamefont{N.}~\bibnamefont{Seiberg}},
  \bibinfo{journal}{Phys. Rev.} \textbf{\bibinfo{volume}{D50}},
  \bibinfo{pages}{1092} (\bibinfo{year}{1994}), \eprint{hep-th/9403198}.

\bibitem[{\citenamefont{Castano et~al.}(1994)\citenamefont{Castano, Piard, and
  Ramond}}]{Castano:1993ri}
\bibinfo{author}{\bibfnamefont{D.~J.} \bibnamefont{Castano}},
  \bibinfo{author}{\bibfnamefont{E.~J.} \bibnamefont{Piard}}, \bibnamefont{and}
  \bibinfo{author}{\bibfnamefont{P.}~\bibnamefont{Ramond}},
  \bibinfo{journal}{Phys. Rev.} \textbf{\bibinfo{volume}{D49}},
  \bibinfo{pages}{4882} (\bibinfo{year}{1994}), \eprint{hep-ph/9308335}.

\bibitem[{\citenamefont{Abbiendi et~al.}(2003)}]{Abbiendi:2003yd}
\bibinfo{author}{\bibfnamefont{G.}~\bibnamefont{Abbiendi}} \bibnamefont{et~al.}
  (\bibinfo{collaboration}{OPAL}), \bibinfo{journal}{Phys. Lett.}
  \textbf{\bibinfo{volume}{B572}}, \bibinfo{pages}{8} (\bibinfo{year}{2003}),
  \eprint{hep-ex/0305031}.

\bibitem[{\citenamefont{Deser and Zumino}(1977)}]{Deser:1977uq}
\bibinfo{author}{\bibfnamefont{S.}~\bibnamefont{Deser}} \bibnamefont{and}
  \bibinfo{author}{\bibfnamefont{B.}~\bibnamefont{Zumino}},
  \bibinfo{journal}{Phys. Rev. Lett.} \textbf{\bibinfo{volume}{38}},
  \bibinfo{pages}{1433} (\bibinfo{year}{1977}).

\bibitem[{\citenamefont{Cremmer et~al.}(1978)}]{Cremmer:1978iv}
\bibinfo{author}{\bibfnamefont{E.}~\bibnamefont{Cremmer}} \bibnamefont{et~al.},
  \bibinfo{journal}{Phys. Lett.} \textbf{\bibinfo{volume}{B79}},
  \bibinfo{pages}{231} (\bibinfo{year}{1978}).

\bibitem[{\citenamefont{Pospelov}(2007)}]{Pospelov:2006sc}
\bibinfo{author}{\bibfnamefont{M.}~\bibnamefont{Pospelov}},
  \bibinfo{journal}{Phys. Rev. Lett.} \textbf{\bibinfo{volume}{98}},
  \bibinfo{pages}{231301} (\bibinfo{year}{2007}), \eprint{hep-ph/0605215}.

\bibitem[{\citenamefont{Cyburt et~al.}(2006)\citenamefont{Cyburt, Ellis,
  Fields, Olive, and Spanos}}]{Cyburt:2006uv}
\bibinfo{author}{\bibfnamefont{R.~H.} \bibnamefont{Cyburt}},
  \bibinfo{author}{\bibfnamefont{J.~R.} \bibnamefont{Ellis}},
  \bibinfo{author}{\bibfnamefont{B.~D.} \bibnamefont{Fields}},
  \bibinfo{author}{\bibfnamefont{K.~A.} \bibnamefont{Olive}}, \bibnamefont{and}
  \bibinfo{author}{\bibfnamefont{V.~C.} \bibnamefont{Spanos}},
  \bibinfo{journal}{JCAP} \textbf{\bibinfo{volume}{0611}}, \bibinfo{pages}{014}
  (\bibinfo{year}{2006}), \eprint{astro-ph/0608562}.

\bibitem[{\citenamefont{Kohri and Takayama}(2007)}]{Kohri:2006cn}
\bibinfo{author}{\bibfnamefont{K.}~\bibnamefont{Kohri}} \bibnamefont{and}
  \bibinfo{author}{\bibfnamefont{F.}~\bibnamefont{Takayama}},
  \bibinfo{journal}{Phys. Rev.} \textbf{\bibinfo{volume}{D76}},
  \bibinfo{pages}{063507} (\bibinfo{year}{2007}), \eprint{hep-ph/0605243}.

\bibitem[{\citenamefont{Kaplinghat and Rajaraman}(2006)}]{Kaplinghat:2006qr}
\bibinfo{author}{\bibfnamefont{M.}~\bibnamefont{Kaplinghat}} \bibnamefont{and}
  \bibinfo{author}{\bibfnamefont{A.}~\bibnamefont{Rajaraman}},
  \bibinfo{journal}{Phys. Rev.} \textbf{\bibinfo{volume}{D74}},
  \bibinfo{pages}{103004} (\bibinfo{year}{2006}), \eprint{astro-ph/0606209}.

\bibitem[{\citenamefont{Ratz et~al.}(2008)\citenamefont{Ratz, Schmidt-Hoberg,
  and Winkler}}]{Ratz:2008qh}
\bibinfo{author}{\bibfnamefont{M.}~\bibnamefont{Ratz}},
  \bibinfo{author}{\bibfnamefont{K.}~\bibnamefont{Schmidt-Hoberg}},
  \bibnamefont{and} \bibinfo{author}{\bibfnamefont{M.~W.}
  \bibnamefont{Winkler}}, \bibinfo{journal}{JCAP}
  \textbf{\bibinfo{volume}{0810}}, \bibinfo{pages}{026} (\bibinfo{year}{2008}),
  \eprint{0808.0829}.

\bibitem[{\citenamefont{Bolz et~al.}(2001)\citenamefont{Bolz, Brandenburg, and
  Buchmuller}}]{Bolz:2000fu}
\bibinfo{author}{\bibfnamefont{M.}~\bibnamefont{Bolz}},
  \bibinfo{author}{\bibfnamefont{A.}~\bibnamefont{Brandenburg}},
  \bibnamefont{and}
  \bibinfo{author}{\bibfnamefont{W.}~\bibnamefont{Buchmuller}},
  \bibinfo{journal}{Nucl. Phys.} \textbf{\bibinfo{volume}{B606}},
  \bibinfo{pages}{518} (\bibinfo{year}{2001}), \eprint{hep-ph/0012052}.

\bibitem[{\citenamefont{Pradler and Steffen}(2007)}]{Pradler:2006qh}
\bibinfo{author}{\bibfnamefont{J.}~\bibnamefont{Pradler}} \bibnamefont{and}
  \bibinfo{author}{\bibfnamefont{F.~D.} \bibnamefont{Steffen}},
  \bibinfo{journal}{Phys. Rev.} \textbf{\bibinfo{volume}{D75}},
  \bibinfo{pages}{023509} (\bibinfo{year}{2007}), \eprint{hep-ph/0608344}.

\bibitem[{\citenamefont{Steffen}(2006)}]{Steffen:2006hw}
\bibinfo{author}{\bibfnamefont{F.~D.} \bibnamefont{Steffen}},
  \bibinfo{journal}{JCAP} \textbf{\bibinfo{volume}{0609}}, \bibinfo{pages}{001}
  (\bibinfo{year}{2006}), \eprint{hep-ph/0605306}.

\bibitem[{\citenamefont{Hashimoto et~al.}(1998)\citenamefont{Hashimoto, Izawa,
  Yamaguchi, and Yanagida}}]{Hashimoto:1998mu}
\bibinfo{author}{\bibfnamefont{M.}~\bibnamefont{Hashimoto}},
  \bibinfo{author}{\bibfnamefont{K.~I.} \bibnamefont{Izawa}},
  \bibinfo{author}{\bibfnamefont{M.}~\bibnamefont{Yamaguchi}},
  \bibnamefont{and} \bibinfo{author}{\bibfnamefont{T.}~\bibnamefont{Yanagida}},
  \bibinfo{journal}{Prog. Theor. Phys.} \textbf{\bibinfo{volume}{100}},
  \bibinfo{pages}{395} (\bibinfo{year}{1998}), \eprint{hep-ph/9804411}.

\bibitem[{\citenamefont{Kallosh et~al.}(2000)\citenamefont{Kallosh, Kofman,
  Linde, and Van~Proeyen}}]{Kallosh:1999jj}
\bibinfo{author}{\bibfnamefont{R.}~\bibnamefont{Kallosh}},
  \bibinfo{author}{\bibfnamefont{L.}~\bibnamefont{Kofman}},
  \bibinfo{author}{\bibfnamefont{A.~D.} \bibnamefont{Linde}}, \bibnamefont{and}
  \bibinfo{author}{\bibfnamefont{A.}~\bibnamefont{Van~Proeyen}},
  \bibinfo{journal}{Phys. Rev.} \textbf{\bibinfo{volume}{D61}},
  \bibinfo{pages}{103503} (\bibinfo{year}{2000}), \eprint{hep-th/9907124}.

\bibitem[{\citenamefont{Felder et~al.}(1999)\citenamefont{Felder, Kofman, and
  Linde}}]{Felder:1999pv}
\bibinfo{author}{\bibfnamefont{G.~N.} \bibnamefont{Felder}},
  \bibinfo{author}{\bibfnamefont{L.}~\bibnamefont{Kofman}}, \bibnamefont{and}
  \bibinfo{author}{\bibfnamefont{A.~D.} \bibnamefont{Linde}},
  \bibinfo{journal}{Phys. Rev.} \textbf{\bibinfo{volume}{D60}},
  \bibinfo{pages}{103505} (\bibinfo{year}{1999}), \eprint{hep-ph/9903350}.

\bibitem[{\citenamefont{Maroto and Mazumdar}(2000)}]{Maroto:1999ch}
\bibinfo{author}{\bibfnamefont{A.~L.} \bibnamefont{Maroto}} \bibnamefont{and}
  \bibinfo{author}{\bibfnamefont{A.}~\bibnamefont{Mazumdar}},
  \bibinfo{journal}{Phys. Rev. Lett.} \textbf{\bibinfo{volume}{84}},
  \bibinfo{pages}{1655} (\bibinfo{year}{2000}), \eprint{hep-ph/9904206}.

\bibitem[{\citenamefont{Giudice et~al.}(1999)\citenamefont{Giudice, Tkachev,
  and Riotto}}]{Giudice:1999yt}
\bibinfo{author}{\bibfnamefont{G.~F.} \bibnamefont{Giudice}},
  \bibinfo{author}{\bibfnamefont{I.}~\bibnamefont{Tkachev}}, \bibnamefont{and}
  \bibinfo{author}{\bibfnamefont{A.}~\bibnamefont{Riotto}},
  \bibinfo{journal}{JHEP} \textbf{\bibinfo{volume}{08}}, \bibinfo{pages}{009}
  (\bibinfo{year}{1999}), \eprint{hep-ph/9907510}.

\bibitem[{\citenamefont{Dine et~al.}(1984)\citenamefont{Dine, Fischler, and
  Nemeschansky}}]{Dine:1983ys}
\bibinfo{author}{\bibfnamefont{M.}~\bibnamefont{Dine}},
  \bibinfo{author}{\bibfnamefont{W.}~\bibnamefont{Fischler}}, \bibnamefont{and}
  \bibinfo{author}{\bibfnamefont{D.}~\bibnamefont{Nemeschansky}},
  \bibinfo{journal}{Phys. Lett.} \textbf{\bibinfo{volume}{B136}},
  \bibinfo{pages}{169} (\bibinfo{year}{1984}).

\bibitem[{\citenamefont{Coughlan et~al.}(1984)\citenamefont{Coughlan, Holman,
  Ramond, and Ross}}]{Coughlan:1984yk}
\bibinfo{author}{\bibfnamefont{G.~D.} \bibnamefont{Coughlan}},
  \bibinfo{author}{\bibfnamefont{R.}~\bibnamefont{Holman}},
  \bibinfo{author}{\bibfnamefont{P.}~\bibnamefont{Ramond}}, \bibnamefont{and}
  \bibinfo{author}{\bibfnamefont{G.~G.} \bibnamefont{Ross}},
  \bibinfo{journal}{Phys. Lett.} \textbf{\bibinfo{volume}{B140}},
  \bibinfo{pages}{44} (\bibinfo{year}{1984}).

\bibitem[{\citenamefont{Joichi and Yamaguchi}(1995)}]{Joichi:1994ce}
\bibinfo{author}{\bibfnamefont{I.}~\bibnamefont{Joichi}} \bibnamefont{and}
  \bibinfo{author}{\bibfnamefont{M.}~\bibnamefont{Yamaguchi}},
  \bibinfo{journal}{Phys. Lett.} \textbf{\bibinfo{volume}{B342}},
  \bibinfo{pages}{111} (\bibinfo{year}{1995}), \eprint{hep-ph/9409266}.

\bibitem[{\citenamefont{Heckman et~al.}(2008)\citenamefont{Heckman, Tavanfar,
  and Vafa}}]{Heckman:2008jy}
\bibinfo{author}{\bibfnamefont{J.~J.} \bibnamefont{Heckman}},
  \bibinfo{author}{\bibfnamefont{A.}~\bibnamefont{Tavanfar}}, \bibnamefont{and}
  \bibinfo{author}{\bibfnamefont{C.}~\bibnamefont{Vafa}}
  (\bibinfo{year}{2008}), \eprint{0812.3155}.

\bibitem[{\citenamefont{Feng et~al.}(2003{\natexlab{a}})\citenamefont{Feng,
  Rajaraman, and Takayama}}]{Feng:2003xh}
\bibinfo{author}{\bibfnamefont{J.~L.} \bibnamefont{Feng}},
  \bibinfo{author}{\bibfnamefont{A.}~\bibnamefont{Rajaraman}},
  \bibnamefont{and} \bibinfo{author}{\bibfnamefont{F.}~\bibnamefont{Takayama}},
  \bibinfo{journal}{Phys. Rev. Lett.} \textbf{\bibinfo{volume}{91}},
  \bibinfo{pages}{011302} (\bibinfo{year}{2003}{\natexlab{a}}),
  \eprint{hep-ph/0302215}.

\bibitem[{\citenamefont{Feng et~al.}(2003{\natexlab{b}})\citenamefont{Feng,
  Rajaraman, and Takayama}}]{Feng:2003uy}
\bibinfo{author}{\bibfnamefont{J.~L.} \bibnamefont{Feng}},
  \bibinfo{author}{\bibfnamefont{A.}~\bibnamefont{Rajaraman}},
  \bibnamefont{and} \bibinfo{author}{\bibfnamefont{F.}~\bibnamefont{Takayama}},
  \bibinfo{journal}{Phys. Rev.} \textbf{\bibinfo{volume}{D68}},
  \bibinfo{pages}{063504} (\bibinfo{year}{2003}{\natexlab{b}}),
  \eprint{hep-ph/0306024}.

\bibitem[{\citenamefont{Borgani et~al.}(1996)\citenamefont{Borgani, Masiero,
  and Yamaguchi}}]{Borgani:1996ag}
\bibinfo{author}{\bibfnamefont{S.}~\bibnamefont{Borgani}},
  \bibinfo{author}{\bibfnamefont{A.}~\bibnamefont{Masiero}}, \bibnamefont{and}
  \bibinfo{author}{\bibfnamefont{M.}~\bibnamefont{Yamaguchi}},
  \bibinfo{journal}{Phys. Lett.} \textbf{\bibinfo{volume}{B386}},
  \bibinfo{pages}{189} (\bibinfo{year}{1996}), \eprint{hep-ph/9605222}.

\bibitem[{\citenamefont{Asaka et~al.}(2000)\citenamefont{Asaka, Hamaguchi, and
  Suzuki}}]{Asaka:2000zh}
\bibinfo{author}{\bibfnamefont{T.}~\bibnamefont{Asaka}},
  \bibinfo{author}{\bibfnamefont{K.}~\bibnamefont{Hamaguchi}},
  \bibnamefont{and} \bibinfo{author}{\bibfnamefont{K.}~\bibnamefont{Suzuki}},
  \bibinfo{journal}{Phys. Lett.} \textbf{\bibinfo{volume}{B490}},
  \bibinfo{pages}{136} (\bibinfo{year}{2000}), \eprint{hep-ph/0005136}.

\bibitem[{\citenamefont{Steffen}(2007{\natexlab{a}})}]{Steffen:2006wx}
\bibinfo{author}{\bibfnamefont{F.~D.} \bibnamefont{Steffen}},
  \bibinfo{journal}{AIP Conf. Proc.} \textbf{\bibinfo{volume}{903}},
  \bibinfo{pages}{595} (\bibinfo{year}{2007}{\natexlab{a}}),
  \eprint{hep-ph/0611027}.

\bibitem[{\citenamefont{Steffen}(2007{\natexlab{b}})}]{Steffen:2007sp}
\bibinfo{author}{\bibfnamefont{F.~D.} \bibnamefont{Steffen}}
  (\bibinfo{year}{2007}{\natexlab{b}}), \eprint{0711.1240}.

\bibitem[{\citenamefont{Affleck and Dine}(1985)}]{Affleck:1984fy}
\bibinfo{author}{\bibfnamefont{I.}~\bibnamefont{Affleck}} \bibnamefont{and}
  \bibinfo{author}{\bibfnamefont{M.}~\bibnamefont{Dine}},
  \bibinfo{journal}{Nucl. Phys.} \textbf{\bibinfo{volume}{B249}},
  \bibinfo{pages}{361} (\bibinfo{year}{1985}).

\end{thebibliography}

\end{document}